# Mapping AI Avant-Gardes in Time:
# Posthumanism, Transhumanism, Genhumanism

**James Brusseau**

Philosophy Department, Pace University, New York City &

Department of Information Engineering and Computer Science, University of Trento,

Italy

## Abstract

Three directions for the AI avant-garde are sketched against the background of time. Posthumanism changes what we are, and belongs to the radical future. Transhumanism changes how we are, and corresponds with the radical past. Genhumanism changes who we are, and exists in the radical present. While developing the concepts, this essay intersects in two ways with theoretical debates about humanism in the face of technological advance. First, it describes how temporal divisions may cleanly differentiate post- and transhumanism. Second, the essay introduces generative humanism, which contributes to discussions about AI and society by delineating a novel humanistic response to contemporary technology. Finally, grounds are provided for a practical project, one where philosophers work with AI engineers in the area of genhumanism. Contemporary AI research into serendipity in recommendation engines provides natural support for the shared research.

## 1. The source of the AI avant-garde is acceleration ethics

The avant-garde in artificial intelligence is generated by acceleration ethics. The essential element of acceleration ethics is the conversion of every innovation into a reason for still faster advance. Naturally, successful innovation reinforces itself, but even when outcomes are harmful, the acceleration continues because of how remedies are sought: through still more innovation [1, 2, 3]. So, when generative AI image platforms are exploited to produce misinformation or deepfake pornography, the response is not to limit AI power or applications [4], instead, it is to expand them, to catalyze research into the automatic detection of harmful content so that it may be deleted [5]. Or, the acceleration response is to develop mathematical distortions to wrap around original images and coat them with perturbations that effectively prevent abusive uses [6]. Innovation itself, in other words, surges forward to resolve innovation's harms.

The same method applies in the area of personal information where applications in AI finance and healthcare have yielded waves of privacy breaches [7]. The acceleration answer is not to restrict data use with prohibitions like the General Data Protection Regulation. Instead, creative algorithmic methods are engineered to increase users' effective control over the release of their information [8]. Or, advances are sought in differential privacy, in the injection of noise into datasets that obscure personal information while maintaining data utility [9]. Either way, the same innovative force that created privacy vulnerabilities leaps ahead to develop new methods of protection. The answer is never less AI, but always more, and faster.

Finally, there is one kind of ethics that addresses harmful technology by diminishing its power [10], and by restricting its use [11], and by pausing its development [12]. Acceleration goes the other way. Advances stimulate more advancing [13], and that logic inevitably leads to the AI avant-garde, it even *is* avant-garde because development is driven by its own logic to ceaselessly lean forward and get ahead of itself.

The subsequent question is: *How?* What human strategies exist on the leading edge once the decision is made to respond to innovation problems with more innovation?

As confined to the human experience of artificial intelligence technology, the purpose of this essay is to respond with a specific analyzing method: the possibilities are distinguished in terms of time. A temporal conception is ontologically first, and it conditions the emergent descriptions of the human avant-gardes. What will emerge are three kinds of time – the radical future, the radical past, and the present of the present – that will yield insights into what posthumanism and transhumanism mean while also, and more significantly, opening a field of provisional explorations of genhumanism.

## 2.  Posthumanism

**Posthumanism** accelerates innovation to change *what* we are. AI and technology convert us into something other than human.

**Vision**: Posthumanism starts with bodies and minds extending past biological limits. This is exoskeletons [14] and augmented reality goggles powered by artificial intelligence. Posthumanism culminates – theoretically – with our minds abstracted from our brains and uploaded to silicon [15, 16].

While the uploading is theoretical, it implies a real victim: me. For traditional humanism, just as the contours of my skin create a space separating me from others in the physical world, so too *my* memories, fears, and aspirations delineate who I am as distinct from everyone else in social reality. So, the double reason I am me and not you is that I am composed from a single physical body with unique psychological states. But, these identifying lines are already blurring at the intersection of extended reality [17] and immersive systems [18]. When interactive headsets and bodysuits connect with  human encounters, these virtual experiences become reproducible. As digitized and automated, whatever happens to me can be played again, and played for anyone. We can all share identical life moments. In fact, we are already beginning to hear hints of this in the new field of research dedicated to the internet of musical things, which uses abundant haptic devices and low-latency digital connections to homogenize concert performances across participants' bodily senses and psychological musical intentions [19]. As the experience of music becomes more enveloping, it is decreasingly subjective and increasingly transferable.

Of course, today's technology offers only narrow, unsatisfying moments, but as the science deepens, the sensations I palpably feel and perceive, and the psychological states I enjoy or endure will increasingly resemble snippets of code that can be cut and pasted from one program – or person – to another. It will get harder, consequently, to hold onto a unique "me." It will get harder to be any *one* person impenetrably different from everyone else.

While it is difficult to imagine how shared and digitally identical experiences could entirely efface the singular moments and aspects of an individual's life, it remains true that individuality splitting across a web of decreasingly unique experiences also threatens to collapse a traditional hierarchy. The original humanist confidence that you and I stand *above* technology falters because we are no longer better than machines, or worse. Instead, we are literally part of them [20].

No matter how farfetched posthuman visions can seem, there is no denying that we are creeping closer to machinic reality: our eyes are learning to focus on distant objects crossing screens pressed close across our faces, and our fingers are learning the unnatural skill of manipulating handheld arrays of joysticks and buttons. What started as human coders, programmers, controllers, scientists, and inventors ends with humans subjected to our own engineering.

Ultimately, posthumanism is the commonly cited vision of humans melding with technology. But there is also an ethereal dimension where differences between me and other people collapse, which occurs alongside the falling hierarchy placing me above machines.

**Paradigm**: Mind uploading. The idea that mentalities and experiences can be transferred to circuitry is fantastical. Even inferior aspirations to cyborg-reality seem fictional [21]. But, roadmaps have been drawn [22], and if the substance of human being is nothing more than material processes as Dennett [23] proposes, and if the ultimate meaning of human existence is pure information as Floridi [24] believes, then we may eventually get past our selves.

**Time**: Radical future. Posthumanism belongs to the future of the future in the sense that what the word "future" means will shift when inhuman durations usurp history and create a new kind of time. One sign of the coming shift is the incongruity between human and generative AI authoring: text production used to require hours, weeks, and years of drafting and editing, but now writing on the scale Borges imagined happens instantly [25].

Beyond time and velocity, a deeper transformation will come. To the degree that extended and immersive reality enmesh our experiences in data and information, our pasts gain the potential to recur without loss or distortion. It is not a reproduction of the past, instead, the experiences are the same production, again. Even if they come later chronologically, there remains the question about what "later" can mean when the moment is identical, all the way down to its digital origin. In this context, the idea of the "true" past and "true" future becomes tenuous, which does not mean false, instead, neither true nor false. Baudrillard [26] called this hyperreality, but no matter the words, the kind of time that belongs to posthumanism will be discontinuous with the time we biologically embodied humans experience now.

Within the context of AI, and as governed by a temporal analysis, the foundation of posthumanism – and what distinguishes it from other directions of the AI avant-garde – is not only the human-machine melding which is already underway everywhere. It is also and more deeply the pending transformation of what it means to coexist with others and technology in time that is reproducible.

## 3. Transhumanism

**Transhumanism** accelerates innovation to change *how* we are: AI and technology perfect our minds and our bodies to make us better, though not fundamentally different. As opposed to the posthuman escape from flesh and blood, here the carnal is purified to meet the expectations of human reason.

**Vision**: The Enlightenment idealized humanity by subjecting our bodies – our "wetware [27]" – to the rules of rationality [28]. From the seventeenth century origin, this purification has been repeatedly achieved through self-denial: the mind imposes its will on flesh through abstinence. The abstinence is easily recognizable as hedonistic denial, but it can also be intellectual. When the French Encyclopedists hid their identities and some of their writings from public dissemination, part of the motivation was fear of science advancing too rapidly for human oversight and control [29]. The fear was that our minds were going too fast for our bodies to manage.

What is distinct and emerging now is the mind's power to govern the body positively. Instead of denial, there is upgrading. Already we have the quantified-self [30], the method of converting our bodies' movements into numbers for rational manipulation on the way to algorithmic fitness. Next comes mechanical and biological supplements, and eventually the Vitruvian man receives brain implants so that our thinking can keep pace with the computers we have created and the rivers of data we are generating [31]. At every advancing step, our biology escalates toward an ideal defined by mental reason more than bodily compulsion. But, it is still *our* biology, our bodies, our humanity that is escalating.

Perhaps the most graphic vision is erobotics: seduction perfected by machine learning and optimized with robots [32]. While this particular experience remains unconsummated, the conceptual reversal at its core is apparent. Where human passion began as the pursuit of a few shared moments of irrational, bodily ecstasy, it will end with bodies quivering under the orchestration of tuned artificial intelligence.

So, transhumanism means enhancing conventionally embodied human being. We remain what we have been, but perfected.

**Paradigm**: Brain implants. This does not transform what our reason *is*, and it does not divide our thoughts from our bodies. It does not make us posthuman. Instead, what reason *does* is supported by accelerating thinking to digital velocity.

**Time**: Radical past. Transhumanism comes from the past in the chronological sense that many of the humanity-plus [33] visions of today are inspired by handwritten documents from the seventeenth and eighteenth centuries. It is more than that, though. Transhumanism also belongs to the past of the past because the birth of the Enlightenment redefined *why* chronological time existed, it reformed what time was *for*, and consequently created its own beginning of history.

At the critical moment of transition from the Medieval to the Modern era, the orthodoxy of human redemption through divine faith was displaced in Europe by the Enlightenment search for salvation through earthly rationality. Descartes incarnated the transition as his dual commitments to religion and to mathematics, in the shadow of Galileo's experience [34], and the result was that transhumanism's origin came before the past as we understand the word. It came from a past dedicated to proving God's existence, as opposed to *our* shared past where the word proof applies to broad scientific understanding. Transhumanism also emerged, consequently, from a specific and powerful notion of what it means to be human: it means being created – our minds *and* our bodies – in the image of God, and that kind of creation has implications. One is that it makes no sense to go *beyond* the conventional understanding of ourselves as envisioned by posthumanism. At least assuming seventeenth century religion, there is no beyond, which means the idea of posthumanism is dismissed not as scientific fantasy so much as religious heresy.

Within the context of AI, and as governed by a temporal analysis, the foundation of transhumanism – and what distinguishes it from other directions of the AI avant-garde – is its historical origin. It came from a time where the idea of perfecting human being could only mean improving how we are as embodied, rational beings.

**Posthumanism and transhumanism**: At the risk of redundancy, it will be noted that the relation between post- and transhumanism is knotted and contested [35]. Central ideas like mind uploading (which separates the mind and the body) and brain implants (which unite the mind and body with technology) relate broadly to the larger themes. This essay is narrow, it is limited to defining, post- and transhuman horizons along a temporal perspective which 1) does offer the virtue of a clear distinction between the two, and 2) also opens the way to a discussion of a third kind of AI avant-garde, one that is firmly situated in the history of philosophy, and in the contemporary work of AI engineers.

## 4. Genhumanism

**Genhumanism** – generative humanism – accelerates innovation to change *who* we are. AI and technology generate new ways to be *a* human.

**Vision**: Originally, it was geography that trapped people in their own lives and identities. When your opportunities are restricted to a town, an office, a nearby social network, then who you can meet, what you can do, where you can explore, all of it remains local [36].

You *are* where you live. For that reason, the early history of generative humanism is populated by travelers, by those who went abroad not to enrich and expand who they were, but to escape their local identities and become someone else. An exemplary case is Isabelle Eberhardt, the 19th century minor aristocrat who traveled to North Africa and went native by changing her language, religion, diet, and nearly everything that it meant to be her [37]. Connie Converse, Neal Cassidy, Paul Bowles and others have gone this way.

One established theoretical foundation for the practice of going somewhere different to become someone else is Deleuze's work – frequently accompanied by Felix Guattari – on the subject of nomadology. Located especially in the "War Machine" chapter of *Thousand Plateaus* [38], the two authors create a double conception of personal identity. First, and intrinsically, identity is multiple. We are not single people with multiple facets so much as multiple people masquerading as continuously recognizable and homogeneous. Second, identity is a production more than an existence, meaning who I am – how I understand my personal values and projects – is a product of my active interaction with the world around me. Even this articulation is not quite right because it implies a pre-existing me who interacts with the world as opposed to an interaction that subsequently creates the me.

This is not the occasion to belabor the theory, though perhaps Deleuze's most accessible formulation can be found in the appendix to the *The Logic of Sense*, "Plato and the Simulacrum," [39]. Additionally, there is Gilbert Simondon's [40] concept of the "pre-individual," which provides the foundation for the emergence of an identifiable person *from* their experiences and actions, as opposed to positing the individual as the causal force shaping experience and acts. Regardless, what matters in the current context is that what I am *doing* generates who I *am*, more than my identity determining what I do. This dynamic allows nomads to be understood not as those who escape their local reality to be somewhere different, but as those who escape where they are to become someone else.

Nomadism is jeopardized today, however. Rapid transportation, economic McDonaldization [41], and electronic networks are compressing geography, making it harder to find someplace different, and still harder to go somewhere unfamiliar to become someone else. It is also true, however, that artificial intelligence compensates by opening contemporary portals of dual identity: now it is big data processes instead of geographical distances that simultaneously capture who we are while generating escape routes.

The capturing side of data and algorithms is most obtrusive in predictive analytics where current recommendation systems ensure that the music we experience, the professional opportunities we find, the partners we meet, all of it is curated to reinforce our personal habits and preferences. Our identities lock in as our customary satisfactions intensify.

At the same time, however, the technology also opens breakaway possibilities. Some are being explored by digital nomads [42], but the most consequential may involve algorithmic serendipity [43]. Coders are exploring techniques for generating recommendations dissimilar from established tastes, but nevertheless engaging. Whether the suggestion is a film, a professional opportunity, or a person for our social or sentimental lives, the search is for a proposal that is unexpected but natural, foreign but also indigenous. [44, 45]

Inspiration comes from hybrid recommendation frameworks linking differences on one level with similarities on another [46]. For example, if artificial intelligence can be trained to find two vastly different people – perhaps divided by languages, by urban and rural, by decades – and subsequently locate a cluster of overlapping interests between them, then one place to explore for new and generative recommendations would be among those interests not currently shared.

Less humanly intuitive, but equally relevant for machine learning, researchers have twisted language models into tools for the production of differences by inverting the coded purpose of locating and ranking high levels of word co-occurrence. Instead of the orthodox method of seeking a next word or token that would likely follow a given one (down- implies "town" or "hill"), what can be sought and surfaced are previously ignored and highly unlikely possibilities at the spectrum's other end. For one research group focused on dietary language, the result was taste innovations – without any quality guarantee: chocolate and garlic, bananas and basil [47]. Another group [48] adapted the approach to suggest directions for movie recommenders. Overall, results have yet to reach consistently high levels of effectiveness: serendipity means that a proposal is unexpected and also satisfying, and getting both is difficult.   Still, the key is to maximize differences and *then* filter for similarities, as opposed to the easier method of seeking (cosine) similarities while ignoring differences [49].

When that happens – when divergences between how people live convert into opportunities for previously unimaginable living – AI generative humanism begins. The next step is to follow breakaway recommendations into broader, identity-transforming experiences, like those traditional humanists have already explored as travelers who go abroad and go native. One embryonic but suggestive example for the digital world is the case of Max Hawkins, the Google engineer who coded a Facebook hack to scan for and locate nearby public events, and then assign one at random [50]. The results were not particularly unique: he ended up spending a night drinking white Russians at a Russian bar, and exercising through an acrobatics and yoga class, and sharing a pancake breakfast at a community center. Still, the experiences were unique to *him* in the sense that he would not have ended up in these places if left to his own habits and tastes. Obviously, this is a rudimentary case of serendipity because simple randomness is being substituted for the work of discovering options that are both different and the same from those previously appreciated, and also because Hawkins understands that *he* is having random experiences, as opposed to the genhumanist idea that serendipitous experiments do – and create – who the experiencer is. Still, in this area where everything is nascent, there is a spark of potential.

Leaving aside the coding challenge of automatically suggesting something that is both different from the standard recommendation and also the same, the work that remains for the contemporary genhumanist is no different from that already confronted in the deserts of north Africa as chronicled by Isabelle Eberhardt, and along the highways of the western United States as told by Jack Kerouac.

So, on one level, artificial intelligence changes nothing: for those tempted by identity disruption, it remains true that shifts in beliefs, languages, homes, customs, professions, social affiliations, and partners, all potentially reshape *who* someone is. What has changed is underneath, however. It is that data and algorithms are no longer about reinforcing existing preferences and thereby deepening self-understanding, instead, they are about developing unexpected interests to catalyze unforeseeable selves.

**Paradigm**: AI serendipity recommender systems, with the critical qualification that serendipity is not oriented toward the conventional purpose – both psychological and economic – of keeping users predictable. Typical AI serendipity serves corporate ends by responding to common business problems, like users abandoning a recommendation platform because suggestions become repetitive or boring. The response is to offer just enough change to maintain interest, but not so much that it scrambles user profiles and so decreases their economic value as targets of marketing strategies and as subjects of surveillance capitalism [51]. To make the effort pay, the unfamiliarity produced by serendipity must ultimately lead back to customary behaviors [52]. To be profitable, users must be predictable.

The kind of serendipity engine that is paradigmatic here is different. Genhumanism serendipity leads *away* from everything familiar, and away from the ideal of familiarity. It is serendipity *for* the differences which attract through ephemeral similarities, as opposed to the conventional aim of finding similarities that are attractive because they are superficially different.

**Time**: Radical present. The ordinary present transitions the past into the future through congruence: it is because one moment *resembles* the next that you cross obliviously between [53]. The radical present – the time of genhumanism – is different, it is the present of the present in the sense that it performs the function of now: it *splits* the past and the future. The commonest human experiences may be splitting away from a consuming job or a lover. The moment is disjunctively present because it so viscerally divides the past's memories and sentiments from the future's open opportunities: it is life stretched between the irreconcilable before and after. So, if genhumanism is the imperative to become someone else, then the only time when that can occur is in a divided but single moment.

The contrary time – the mirror image – might be Heidegger's conception of making present in *Being and Time* [54] which reverses the genhuman present in two ways. First, for Heidegger, the present is *about* the interpolation of past and future. The only question is: In what way are they funneled through now? Heidegger's response is well known. The task of the present for human being is to account for our personal past while projecting forward into our own (or, our own-most) destiny. The present is the *way* we do that. For genhumansim, by contrast, it is the reverse: the present is the *way* that a future can exist precisely as *not* accounting for the past, as escaping it. Then, secondly, for Heidegger the ethical challenge of the present is to be resolute, it is to take responsibility for our own past while orienting toward a future which is authentically our own, one that reflects the projects defining our true place in the world. For genhumansim, by contrast, it is the reverse: the ethical challenge of the present is to take responsibility for a future that will be established independently of – and that holds no responsibility for – the past.

All this is theoretical, of course. Barring cerebral injury, no one simply exorcises their memories, fears, aspirations, and values. No one entirely walks away from their own past. But, just like the serendipitous recommendation of a song on a music platform, or of a restaurant on a dining website, it is possible to be engrossed in an aspect of the future that was unforeseeable. And, it is possible to leverage data and algorithms to push further, to experiment with an identity that is generated at a moment in future-oriented time more than developed through past time.

Finally, the temporal foundation of genhumanism – and what distinguishes it from other directions of the AI avant-garde – is that it is based on human differences through time, which does not mean differences between humans, but within each one, and as the origin of each one. Posthumanism, contrastingly, creates a difference in time for humans, but on the way to the inhuman. And, transhumanism projects humanity without differences in time when progress allows diverse human-beings to converge as the Enlightenment ideal.

**Avant-garde genhumanism and working engineers**: One invigorating element of genhumanism is that AI engineers are working urgently on the challenge of triggering serendipity today, in both corporate and university labs. Part of their motivation is the intersection with profit, the possibility of providing users with just enough of the unexpected to keep them engaged while not triggering an escape into reformist interests. While it is true that profit-driven research can be understood this way, as leveraging the unpredictable to ultimately reinforce predictable personal identities that satisfy marketers seeking foreseeable behaviors [55, 56], it is also true that advances in serendipity knowledge cannot be siloed by economic interests. There will always be the potential for engineers and users to twist established discoveries for unorthodox purposes. And, to the degree those purposes coincide with genhumanism, that particular avant-garde steps away from the other two. It does because this form of humanism can be constructed on a foundation of existing, widespread technical research.

One hesitation surrounding posthumanism and, to a lesser extent, transhumanism is the suspicion that the relative absence of real-world engineering dedicated to the projects parallels a lack of philosophical and analytic rigor in defining exactly what the projects *are* [57, 58]. To the extent few practicing engineers are actually pursuing post- and transhumanism, theorists are left with excessive room for speculation. Because mind uploading, for instance, does not even remotely exist, accompanying abstract ideas can seem more like science fiction than social science, and while this is not the place to enter those debates, what does fit here is a description of why genhumanism may escape this kind of criticism.

It may escape because real-world engineering *is* happening, and pervasively. Detailed taxonomies already exist to understand the serendipity problem in the way data scientists do, by cutting experience into discrete units that can be separately understood and subsequently compiled. Just like there are routines and sub-routines, so too there are questions and sub-questions. For serendipity, the sub-questions ask about: unexpectedness [59], novelty [60], and diversity [61]. Then, the research itself has already been channeled toward discrete responsibilities – pre-processing, in-processing, and post-processing – each with its own priorities, and separate accomplishments and failures [62]. There are publicly available datasets for experimenting and testing, including sets from Amazon [63], Yelp [64], and Netflix [65]. There are established obstacles for deep learning and reinforcement learning models: "ground truth data generation for serendipity," "timing for serendipity recommendations," "cross-domain learning for serendipity recommendations" [66].

More could be added, but the technical details are secondary. What matters first is the implication. The fields of post- and transhumanism have been derided as academically speculative and unserious, which may or may not be true. It is definitely true, however, that genhumanism entwines with a long strand of venerable philosophical history, and it is built on detailed artificial intelligence research that is underway and well-funded.

## 5. Further analysis

The foreground research problem addressed in this essay is the origin and meaning of genhumanism as an aspect of the AI avant-garde. It emerged from the author's experience in lecturing across disciplines, in a Department of Philosophy, and also in a Department of Information Engineering and Science. That background provided both the conceptual and technical ingredients of this paper.

Then, in the paper's background, there stands a more established research problem: What is the differences between post- and transhumanism? Because the two frameworks for meeting technological advance overlap in theory and in lived experience, debates have emerged about their defining features [67].

The conflicts are not only products of direct scholarly disagreement but also differences in scholarly backgrounds. Questions surrounding artificial intelligence and technological advance have been addressed by philosophers from both the continental [68] and analytic [69] traditions, by theologians 70], by computer scientists, and engineers. Literary theorists, artists, and writers have also proposed conceptions and approaches [71]. Given the perspectival abundance, it is unsurprising to find divergent understandings.

Through the uncertainties, there cuts a persistent conceptual distinction. It clefts the posthumanism belonging to the history of advancing technology, from the posthumanism belonging to the history of European philosophy, and its frustratingly tangled accomplishments and setbacks.

1) When embedded in technological advance, posthumanism can be conceived as an extension or outcome of transhumanism. After humans become perfect, innovation keeps pushing beyond to more than perfect, to the inhuman, to the beyond human. Nick Bostrom [72] describes this posthumanist view cogently in his Transhumanist FAQ.

2) The idea of posthumanism *also* traces back through continental philosophy, including Deleuze and Guattari's 1970s collaborations [73]. Here, the core idea is not that humans *advance* to a superior posthuman phase, instead, we *recede* into our own surrounding environment. What is questioned, fundamentally, is the intrinsic privilege of human agency [74]. Our intentions and projects, according to this post-anthropocentric view, are no more ontologically profound than the enveloping natural world. So, as opposed to Aquinas's natural law consigning plant and animal life to the service of humans, 1970s European posthumanism compresses all nature onto the same level of intrinsic significance [75]. Without the advantage of hierarchical privilege, we humans can no longer command the assembly of materials and ambitions in the world, instead, we are cogs and elements in larger machines and machinic desires [76].

In essence, the technologically oriented posthumanists conceive their vision as a triumph and advance, whereas the older, continental philosophical approach defines the word as something nearer a humbling of humanity. This genetic difference reverberates back through the entire discussion of the AI avant-gardes as a question: Are they triumphs, or something closer to humiliations?

Answering belongs to a different essay.

What has been constructed here leads to a narrower argument. It starts from the observed confusion between the various humanisms surrounding contemporary technology and artificial intelligence. Then it proposes that their conceptual divisions may be elucidated by appealing to time. From there, a clean split emerges between post- and transhumanism, while a novel field of inquiry – genhumanism – organically spreads open.

## 6. Conclusion

Acceleration ethics thrusts humanism onto the leading edge of always intensifying innovation where there are three ways of belonging in time, and three human strategies for converging with artificial intelligence. They are represented in *Figure 1*, and it remains uncertain whether today's avant-garde will seek to be other, perfect, or different.

Finally, this paper has offered two theoretical contributions. It adds clarity to the posthuman and transhuman distinction, while also nurturing the concept of generative humanism. Then, for those interested in the avant-garde as something to *do* more than understand, there is also a practical implication: the fostering of collaborations between philosophers and AI engineers.

For philosophers, collaborating will require an effort to understand the mathematical and statistical languages of similarities and differences. For example, there is the meaning and use of cosine similarities by engineers experimenting with recommendation engines. Then, for engineers, they will need to understand what difference means on the conceptual level, not as inferior degrees of similarity but as kinds of otherness. For example, there is the idea that current work in serendipity could be twisted to function outside its conventional purpose. Instead of users being offered unfamiliar options as a method of maintaining engagement and ensuring that future recommendations are satisfying, the algorithmic engines can be warped to create ways out, to provide options that allow users to find their own ways to becoming unpredictable.

|  | Change | Role of AI | Time |
|---|---|---|---|
| **Posthumanism** | What we are | Convert us into something other than human | Future of the future, after the end of humanism |
| **Transhumanism** | How we are | Perfect us as humans | Past of the past, at the birth of the Enlightenment's ideal human |
| **Genhumanism** | Who we are | Generate new ways to be a human | Present of the present, as splitting the past and future of individual human lives |

Figure 1. Posthumanism, Transhumanism, Genhumanism

## Declaration

Author declares there is no conflict of interest or external funding.

## References

[1] Brusseau, J. Acceleration AI Ethics, the Debate between Innovation and Safety, and Stability AI's Diffusion versus OpenAI's Dall-E . *International Conference on Computer Ethics: Philosophical Enquiry (CEPE 2023)*. Chicago, USA. 2023; https://journals.library.iit.edu/index.php/CEPE2023/article/view/293.

[2] Fuller, S. Technological Unemployment as a Test of the Added Value of Being Human. In: Peters, M., Jandrić, P., Means, A. (eds) Education and Technological Unemployment. Springer, Singapore. 2019; https://doi.org/10.1007/978-981-13-6225-5_8.

[3] Sunstein, C. The precautionary principle as a basis for decision making. The Economists' Voice 2, no. 2. 2005. https://dash.harvard.edu/bitstream/handle/1/29998410/PrecautionaryPrinciple.pdf Accessed 8 September 2023.

[4] Yudkowsky E. Pausing AI Developments Isn't Enough. We Need to Shut it All Down. Time Magazine: March 29. 2023. https://time.com/6266923/ai-eliezer-yudkowsky-open-letter-not-enough/. Accessed 8 September 2023.

[5] Rana M, Nobi M, Murali B, Sung A. Deepfake detection: A systematic literature review. IEEE Access; 2022.

[6] Salman H, Alaa K, Leclerc G, Ilyas A, Madry A. Raising the Cost of Malicious AI-Powered Image Editing. Gradient science. 2022; https://gradientscience.org/photoguard/

[7] Singh M, Halgamuge M, Ekici G, Jayasekara C. A Review on Security and Privacy Challenges of Big Data. In: Sangaiah A, Thangavelu A, Meenakshi Sundaram V. (eds) Cognitive Computing for Big Data Systems Over IoT. Lecture Notes on Data Engineering and Communications Technologies, vol 14 . Springer, Cham. 2018; https://doi.org/10.1007/978-3-319-70688-7_8

[8] Cummings R, Hadi E, Gkatzelis V, Pountourakis E, Ziani J. Optimal data acquisition with privacy-aware agents." arXiv preprint. 2022; arXiv:2209.06340.

[9] Zheng H, Hu H, Han Z, Preserving User Privacy for Machine Learning: Local Differential Privacy or Federated Machine Learning? In *IEEE Intelligent Systems*, vol. 35, no. 4, pp. 5-14, 1 July-Aug. 2020; https://doi.org/10.1109/MIS.2020.3010335.

[10] Yudkowsky E. Pausing AI Developments Isn't Enough. We Need to Shut it All Down. Time Magazine: March 29. 2023. https://time.com/6266923/ai-eliezer-yudkowsky-open-letter-not-enough/. Accessed 8 September 2023.

[11] Leslie, D. Understanding artificial intelligence ethics and safety. arXiv preprint. 2019; arXiv:1906.05684.

[12] Clarke, L. Call for AI pause highlights potential dangers. Science. (New York, NY) 380, no. 6641. 2023; 120-121.

[13] Brusseau, J. How to Accelerate Ethics for Innovation and Against Precaution in Generative AI. *Meeting of the International Association for Computing and Philosophy (IACAP-2023 CEVAST)*. Prague, Czechoslovakia. 2023; https://doi.org/10.5281/zenodo.8024360.

[14] Ottobock. Ottobock bionic exoskeltons. Ottobock Bionic Exoskeletons. 2023; https://ottobockexoskeletons.com/ Accessed 7 September 2023.

[15] Watanabe M. Uploading Human Consciousness. In From Biological to Artificial Consciousness. The Frontiers Collection. Springer, Cham. 2022; https://doi.org/10.1007/978-3-030-91138-6_6

[16] Bostrom, N. Ethical issues in advanced artificial intelligence. Science fiction and philosophy: from time travel to superintelligence 277; 2003.

[17] Pyun K., Rogers, J, Ko, S. Materials and devices for immersive virtual reality. Nat Rev Mater **7**, 841–843. 2022; https://doi.org/10.1038/s41578-022-00501-5.

[18] Adams, D. Virtual Retail in the Metaverse: Customer Behavior Analytics, Extended Reality Technologies, and Immersive Visualization Systems. Linguistic and Philosophical Investigations 21. 73–88. 2022; https://doi.org/10.22381/lpi2120225

[19] Turchet L, Fazekas G, Lagrange M, Ghadikolaei H, Fischione C. The internet of audio things: State of the art, vision, and challenges. IEEE Internet of Things Journal 7, no. 10. 2020; 10233-10249.

[20] Santos, S. Posthuman Knowledge. Journal of Posthuman Studies 4 August; 4 (1): 107–112. 2020; https://doi.org/10.5325/jpoststud.4.1.0107

[21] Park, J. Shakespeare in Cyborg Theatre: Immersive VR Theatre and the Cyborg-Subject, Contemporary Theatre Review, 32:2, 177-190. 2022; https://doi.org/10.1080/10486801.2022.2047031

[22] Sandberg A, Bostrom N. Whole Brain Emulation: A Roadmap, Technical Report #2008-3, Future of Humanity Institute, Oxford University. 2008; www.fhi.ox.ac.uk/reports/2008-3.pdf. Accessed 7 September 2023.

[23] Dennett, D. From Bacteria to Bach and Back: The Evolution of Minds. New York: W. W. Norton & Company; 2018.

[24] Floridi, L. The Informational Nature of Personal Identity Minds & Machines 21, 549–566. 2011; https://doi.org/10.1007/s11023-011-9259-6.

[25] Basile, J. Library of babel. 2023. https://libraryofbabel.info/. Accessed 18 July 2023.

[26] Baudrillard, J. *Simulacra and Simulations.* Ann Arbor: University of Michigan Press; 1994.

[27] Ruby, A. Gendering the Posthuman: The Intersection of Gender, Technology, and Control on the Cyborg Body in Garland's Ex Machina. Aletheia: The Arts and Science Academic Journal Vol. 2 No. 1; 2022.

[28] Descartes, R. Meditations on first philosophy. M. Moriarty, Trans. Oxford: Oxford University Press; 2008.

[29] Rosen, S. The Mask of Enlightenment: Nietzsche's Zarathustra, Second Edition. Yale University Press. 2004: http://www.jstor.org/stable/j.ctt32bnxd. Accessed 18 February 2023.

[30] De Moya, J-F, Pallud, J. From panopticon to heautopticon: A new form of surveillance introduced by quantified-self practices. Inf Syst J.; 30: 940– 976. 2020; https://doi.org/10.1111/isj.12284

[31] Jewett C, Metz C. Elon Musk hopes to test a brain implant in humans next year. The New York Times, December 1. 2022; https://www.nytimes.com/2022/11/30/health/elon-musk-neuralink-brain-device.html. Accessed 7 September 2023.

[32] Dubé S, Anctil D. Foundations of Erobotics." International journal of social robotics vol. 13,6: 1205-1233. 2021; https://doi.org/10.1007/s12369-020-00706-0

[33] Humanity+. March 27, 2023, humanityplus.org  2023; https://www.humanityplus.org/ Accessed 7 September 2023.

[34] Nadler S, Schmaltz T,  Antoine-Mahut D. eds. The Oxford Handbook of Descartes and Cartesianism. Oxford: Oxford University Press; 2019.

[35] Fuller S, Lipińska V. The Proactionary Imperative: A Foundation for Transhumanism. London: Palgrave Macmillan. 2014; https://doi.org/10.1057/9781137302922.

[36] Ash J, Kitchin R, Leszczynski, A. Digital turn, digital geographies? Progress in Human Geography. *42*(1), 25–43. 2018; https://doi.org/10.1177/0309132516664800

[37] Abdel-Jaouad, H. Isabelle Eberhardt: Portrait of the Artist as a Young Nomad. Yale French Studies, no. 83. 93–117. 1993; https://doi.org/10.2307/2930089

[38] Deleuze G, Guattari F. A Thousand Plateaus: Capitalism and Schizophrenia. Minneapolis: University of Minnesota Press; 1987.

[39] Deleuze, G. The Logic of Sense, trans. M. Lester. London: Continuum; 2004.

[40] Simondon, G. Technical Mentality in A. de Boever, A. Murray, J. Roffe, A. Woodward (eds) Gilbert Simondon: Being and Technology. Edinburgh: Edinburgh University Press; 2012.

[41] Ritzer, G, Miles S. The changing nature of consumption and the intensification of McDonaldization in the digital age. Journal of Consumer Culture 19, no. 1. 2019; 3-20.

[42] Brusseau, J. What does it mean to be a digital nomad? *Turkish Policy Quarterly.* 2022. Retrieved May 3, 2023, from http://turkishpolicy.com/article/1104/what-does-it-mean-to-be-a-digital-nomad.

[43] Smets A, Vannieuwenhuyze J, Ballon P. Serendipity in the city: User evaluations of urban recommender systems. Journal of the Association for Information Science and Technology, 73(1), 19– 30. 2021; https://doi.org/10.1002/asi.24552

[44] Boldi R, Lokhandwala A, Annatone E, Schechter Y, Lavrenenko A, Sigrist C. Improving Recommendation System Serendipity Through Lexicase Selection. arXiv preprint 2023; arXiv:2305.11044.

[45] Fu Z, Niu X, Maher M. Deep Learning Models for Serendipity Recommendations: A Survey and New Perspectives. ACM Computing Surveys. 2023; https://dl.acm.org/doi/pdf/10.1145/3605145.

[46] Ponti M, Skarpeti A, Kestemont B. AI and Citizen Science for Serendipity. arXiv preprint. 2022; arXiv:2205.06890.

[47] Grace K, Maher M, Davis N, Eltayeby O. "Surprise walks: Encouraging users towards novel concepts with sequential suggestions." In Proceedings of the 9th International

Conference on Computational Creativity (ICCC 2018). Association of Computational Creativity; 2018.

[48] Li X, Wenjun J, Chen W, Wu J, Wang G. HAES: A new hybrid approach for movie recommendation with elastic serendipity. Proceedings of the 28th ACM International Conference on Information and Knowledge Management. 1503-1512; 2019.

[49] Varghese N, Punithavalli M. Semantic Similarity Extraction on Corpora Using Natural Language Processing Techniques and Text Analytics Algorithms. In: Mathur G, Bundele M, Lalwani M, Paprzycki M. (eds) Proceedings of 2nd International Conference on Artificial Intelligence: Advances and Applications. Algorithms for Intelligent Systems. Springer, Singapore. 2022; https://doi.org/10.1007/978-981-16-6332-1_16.

[50] Spiegel A, Rodríguez M. Eager To Burst His Own Bubble, A Techie Made Apps To Randomize His Life. National Public Radio. 2017.

[51] Zuboff S, Möllers N, Wood D, Lyon D. Surveillance Capitalism: An Interview with Shoshana Zuboff. *Surveillance & Society* 17, no. ½ 2019; 257-266.

[52] Reviglio, U. Serendipity as an emerging design principle of the infosphere: challenges and opportunities. Ethics and Information Technology 21, no. 2. 2019; 151-166.

[53] Smith, S. Temporal Relativism and the Objective Present. Journal of Posthuman Studies 1 June; 5 (1): 39–52. 2021; https://doi.org/10.5325/jpoststud.5.1.0039

[54] Heidegger, M. Being and Time.Translated by John MacQuarrie and Edward Robinson. New York: Harper & Row Publishers; 1962.

[55] Niu X, Al-Doulat A.Lucky Find: leveraging surprise to improve user satisfaction and inspire curiosity in a recommender system. Proceedings of the 2021 Conference on Human Information Interaction and Retrieval. 163-172; 2021.

[56] Deshpande M, Karypis G. Item-based top-n recommendation algorithms. In ACM Transactions on Information Systems (TOIS) 22, no. 1.(2004; 143-177.

[57] Hauskeller, M. Mythologies of Transhumanism. London: Palgrave Macmillan. 2016; https://doi.org/10.1007/978-3-319-39741-2.

[58] Bostrom, N. Why I want to be a posthuman when I grow up. In The transhumanist reader: Classical and contemporary essays on the science, technology, and philosophy of the human future. 2013; 28-53.

[59] Zhang M, Yang Y, Abbas R, Deng K, Li J, Zhang B. SNPR: A Serendipity-Oriented Next POI Recommendation Model. In Proceedings of the 30th ACM International Conference on Information & Knowledge Management. 2021; 2568-2577.

[60] Shrestha P, Zhang M, Liu Y, Ma S. Curiosity-inspired Personalized Recommendation. In 2020 IEEE/WIC/ACM International Joint Conference on Web Intelligence and Intelligent Agent Technology (WI-IAT). 2020; 33-40.

[61] Lee J, Jung S, Chae D. A diversity personalization approach towards recommending POIs. In Proceedings of the 37th ACM/SIGAPP Symposium on Applied Computing. 1804-1807; 2022.

[62] Zhe F, Xi N, Maher M. Deep Learning Models for Serendipity Recommendations: A Survey and New Perspectives. ACM Comput. Surv. 2023; https://doi.org/10.1145/3605145.

[63] Li X, Wenjun J, Weiguang C, Jie W, Guojun W, Li K. Directional and explainable serendipity recommendation. Proceedings of The Web Conference 2020. 122-132; 2020.

[64] Asghar, N. Yelp dataset challenge: review rating prediction. arXiv preprint. 2016; arXiv:1605.05362

[65] Bennett, J. Lanning. S. The Netlix Prize. In Proceedings of KDD cup and workshop, Vol. 2007. New York, NY, USA., 2007; 35.

[66] Zhe F, Xi N, Maher M. Deep Learning Models for Serendipity Recommendations: A Survey and New Perspectives. ACM Comput. Surv. 2023; https://doi.org/10.1145/3605145.

[67] Sümeyra B, Çağdaş D, Pelin K, Tuncel, Y. "Posthumanisms beyond Disciplines," Journal of Posthumanism, Transnational Press London, UK, vol. 1(1), 2021; 1-4, May.

[68] Braidotti, R. Posthuman Critical Theory. Journal of Posthuman Studies, vol. 1, no. 1, pp. 9–25. 2017; https://doi.org/10.5325/jpoststud.1.1.0009

[69] Chalmers, D. The Conscious Mind: In Search of a Fundamental Theory. New York: Oxford University Press; 1996.

[70] Ferrando, F. Leveling the Posthuman Playing Field, Theology and Science, 18:1, 1-6. 2020; https://doi.org/10.1080/14746700.2019.1710343.

[71] Kirsch, Adam. The Revolt Against Humanity: Imagining a Future Without Us. New York: Columbia Global Reports; 2023.

[72] Bostrom, N. Transhumanist FAQ. https://nickbostrom.com/views/transhumanist.pdf. 2003b. Accessed 22 August 2023

[73] Deleuze G, Guattari F. A Thousand Plateaus: Capitalism and Schizophrenia. Minneapolis: University of Minnesota Press; 1987.

[74] Hui Y, Morelle L. A Politics of Intensity: Some Aspects of Acceleration in Simondon and Deleuze. Deleuze Studies 11:4. 2017; 498-517.

[75] Braidotti, R. Posthuman Critical Theory. Journal of Posthuman Studies, vol. 1, no. 1, pp. 9–25. 2017; https://doi.org/10.5325/jpoststud.1.1.0009

[76] Deleuze G, Guattari F. A Thousand Plateaus: Capitalism and Schizophrenia. Minneapolis: University of Minnesota Press; 1987.